\documentclass[%
 reprint,
 amsmath,amssymb,
 aps,
]{revtex4-2}

\usepackage{graphicx}
\usepackage{dcolumn}
\usepackage{bm}
\usepackage{tabularx}
\usepackage{subcaption}
\usepackage{xcolor}
\usepackage{svg}

\begin{document}

\allowdisplaybreaks



\title{Chirality of a $\mathbb{Z}_q$ Model as Orientation-Odd Phase Shifts in Oscillator Networks}

\author{Yi Cheng}
  \email{zss7gw@virginia.edu}
 \altaffiliation{Charles L. Brown Department of Electrical and Computer Engineering, University of Virginia, Charlottesville, Virginia 22904, USA.}

\author{Zongli Lin}
 \email{zl5y@virginia.edu}
\altaffiliation{Corresponding Author;
Charles L. Brown Department of Electrical and Computer Engineering, University of Virginia, Charlottesville, Virginia 22904, USA.
}





\begin{abstract}
Chiral interaction between discrete $\mathbb{Z}_q$ variables, which distinguishes clockwise from counterclockwise steps, arises in settings ranging from statistical physics to combinatorial optimization. When such an interaction is realized by continuous phases, however, its kernel fixes the phase coupling only at the $q$ grid points, so which part of the coupling carries the discrete chirality is not determined. We show that for any realization that matches the discrete values and treats a kernel and its mirror image alike, the kernel is chiral if and only if the coupling has a nonzero odd part, and that neither property of the realization can be dropped. Among the realizations that reproduce all discrete energies and equilibria, the one with the fewest harmonics is unique and satisfies the second property automatically, so that its sine components measure the discrete chirality exactly; a faithful realization with an even coupling, as in existing oscillator Ising and Potts machines, represents only nonchiral kernels. In the induced dynamics this odd part is an orientation-odd phase shift, which changes sign with the orientation of the coupling, in contrast to the Sakaguchi phase lag. In oscillator networks it is realized by direction-dependent delays, and mixed-level circuit simulations show that exchanging the delays reverses the chirality of the locked states.
\end{abstract}

\maketitle


\textit{Introduction.}---Interactions that distinguish clockwise from counterclockwise steps between discrete states arise in many settings. In statistical physics, chiral clock and chiral Potts models describe commensurate--incommensurate transitions of adsorbed monolayers \cite{ostlund1981incommensurate,huse1981simple,huse1984commensurate} and the ordering transitions of Rydberg-atom chains \cite{whitsitt2018quantum,keesling2019quantum}. Combinatorial optimization problems of the same kind include periodic scheduling with time windows on the differences of event times once times are discretized to $q$ slots per period \cite{serafini1989mathematical}, and maximum-likelihood detection of $q$-ary phase-shift-keyed symbols in multiple-antenna channels (MIMO), where complex channel correlations give chiral clock couplings \cite{steingrimsson2003soft,kim2019leveraging}. In each case, $q$-state variables interact pairwise through a kernel $f$ of the difference $s_i-s_j \bmod q$, and this kernel is generally chiral, $f(k)\neq f(q-k)$: it assigns different costs to a step $k$ and to the opposite step $-k$. Networks of coupled oscillators, which represent the state $k$ by a locked phase $2\pi k/q$, provide a native physical encoding of $\mathbb{Z}_q$ variables, and oscillator Ising, clock, and Potts machines built on them all use nonchiral kernel. When such physical systems are used to simulate a discrete model or to solve its optimization problem \cite{mohseni2022ising}, their dynamics should reproduce the physics of the model faithfully \cite{hauke2012can}, including its symmetry properties and not only its energies \cite{halimeh2020reliability,yang2020observation}. The discrete energies are fixed by matching the kernel on the grid, but chirality is not: as we show, it can be misrepresented even when every discrete energy and equilibrium is reproduced exactly.

For a continuous kernel model, its chirality is unambiguous. Mirror reflection, $\theta_i\to-\theta_i$, acts on a pair coupling as $g(x)\to g(-x)$ with $x=\theta_i-\theta_j$, and every periodic $g$ splits uniquely into an even part, which reflection leaves invariant, and an odd part, which it reverses. The odd part is therefore the unique measure of chirality. It is also the natural one, because the same $g$ generates the dynamics: for the chiral $XY$ coupling $\cos(x+\Delta)$ \cite{ostlund1981incommensurate,huse1981simple}, the odd part $-\sin\Delta\sin x$ is the phase shift $\Delta$ in the equations of motion, and chiral reversal is $\Delta\to-\Delta$ \cite{mahler2025programmable}. For a discrete kernel model, in contrast, chirality is ambiguous once the model is realized by continuous phases. The kernel is a set of $q$ numbers and prescribes the coupling only at the grid points $2\pi k/q$; infinitely many periodic functions agree with it there and differ in their odd parts: adding $\sin(q\theta)$ to the coupling, which vanishes at every grid point, leaves all discrete energies unchanged but creates an odd part even for a nonchiral kernel. Which property of the continuous dynamics represents the discrete chirality, and whether the answer is unique, are not solely determined by the discrete model.

Existing work studies either continuous kernels or special discrete kernels. In continuous models, chirality is a coupling phase from the outset, as in oscillatory neural networks \cite{hoppensteadt2000pattern}, laser arrays with synthetic gauge fields \cite{mahler2025programmable}, and the anisotropic chiral $XY$ model \cite{seno1994spin}. For the discrete clock form $\cos(2\pi(s_i-s_j)/q+\Delta)$, chirality is the offset $\Delta$ by construction, as in $q$-state phasor networks \cite{noest1988discrete} and a Josephson-junction chain realizing a chiral $\mathbb{Z}_3$ clock model \cite{wauters2025engineering}; in quantum chiral Potts chains it is carried by the Fourier phases of the bond operator \cite{howes1983quantum,von1985zn,fendley2012parafermionic}. Oscillator-based Ising, clock, and Potts machines \cite{chou2019analog,wang2021solving,ahmed2021probabilistic,moy20221,lo2023ising,wu2025fully,cilasun2025coupled,cheng2025impacts,kalinin2018simulating,kalinin2018networks,sakaguchi2018community,honari2020optical,honari2022combinatorial,roychowdhury2022oscillator,mallick2022computational,cheng2026physics} map discrete Hamiltonians onto phase-energy Lyapunov functions with subharmonic pinning, but with nonchiral kernels. To our knowledge, a correspondence between discrete chirality and the continuous coupling has not been established for arbitrary $\mathbb{Z}_q$ kernels.

Here we answer the question in general: for every realization that matches the discrete values and treats a kernel and its mirror image alike, chirality is exactly the odd part of the continuous coupling, neither condition can be dropped, and the minimal realization that reproduces all discrete energies and equilibria is unique and has this property. In the phase dynamics, this odd part appears as phase shifts of opposite sign in the two directions of each link: a lag common to both directions, as in the Sakaguchi model, cannot represent chirality, and reversing the handedness swaps the two shifts without changing the coupling strengths. Oscillator networks realize these shifts by direction-dependent delays, which we test in circuit simulations. Since an even coupling has no odd part, oscillator machines with even couplings must embed a chiral kernel in a larger spin space, and we compare the oscillator-per-spin direct realization with a one-hot embedding in time to solution. We begin with the discrete model and its continuous realizations.


\textit{Discrete model and continuous relaxations.}---We consider $N$ spins $s_i\in\mathbb{Z}_q=\{0,1,\dots,q-1\}$ with the pairwise Hamiltonian
\begin{equation}
H_f(\mathbf{s}) = -\sum_{(i,j)\in\mathcal{E}} J_{ij}\, f(s_i\ominus s_j),
\label{eq:H}
\end{equation}
with $s_i\ominus s_j \equiv (s_i-s_j)\bmod q$, where $\mathcal{E}$ is a set of oriented edges (ordered pairs) of an arbitrary graph, $J_{ij}$ are real couplings, and $f:\mathbb{Z}_q\to\mathbb{R}$ is an arbitrary kernel.
Reversing the orientation of an edge is equivalent to replacing $f$ by its reflection on that edge. The reflected kernel is $f^{\rm R}(k)=f(q\ominus k)$; $f$ is nonchiral if $f=f^{\rm R}$ and chiral otherwise. Reflection of the kernel is equivalent to reversing every spin, $H_{f^{\rm R}}(\mathbf{s})=H_f(-\mathbf{s})$, so a chiral kernel distinguishes clockwise from counterclockwise phase differences. Although Eq. \eqref{eq:H} uses a common kernel for simplicity, the results extend edgewise to edge-dependent kernels \(f_{ij}\). On-site terms can likewise be incorporated by introducing a fixed reference spin $(s_0=0)$.

A continuous realization assigns each state $k$ the phase $x_k=2\pi k/q$ and replaces $\mathbf{s}$ by phases $\boldsymbol{\theta}\in[0,2\pi)^N$ with energy
\begin{equation}
E(\boldsymbol{\theta}) = -K\sum_{(i,j)\in\mathcal{E}} J_{ij}\, g(\theta_i-\theta_j) - \frac{K_{\rm s}}{q}\sum_i \cos(q\theta_i),
\label{eq:E}
\end{equation}
where $K>0$ sets the coupling strength and the pinning term of strength $K_{\rm s}>0$ favors the $q$ grid phases. The pair function $g$ must be $2\pi$-periodic: the phases live on a circle, and the coupling functions obtained by phase reduction of limit-cycle oscillators are periodic by construction \cite{kuramoto1984chemical,nakao2016phase}. Within this class, however, $g$ is constrained by $f$ only through its values at the $q$ grid points. We therefore describe a realization by a rule $\Phi$ that assigns to every kernel $f$ a $2\pi$-periodic, continuously differentiable function $g=\Phi[f]$, and we ask which properties of $\Phi$ make the chirality of $g$ a faithful image of the chirality of $f$. Two properties are natural. Value matching,
\begin{equation}
\Phi[f](x_k)=f(k), \quad k\in\mathbb{Z}_q,
\label{eq:value}
\end{equation}
makes the continuous energy (\ref{eq:E}) reproduce $H_f$ on the grid up to a constant. Reflection equivariance,
\begin{equation}
\Phi[f^{\rm R}](x)=\Phi[f](-x)\quad\text{for all } f,
\label{eq:equiv}
\end{equation}
states that the rule treats a kernel and its mirror image in the same way.

\textit{Theorem 1 (Chirality is the odd part).}---Let $\Phi$ satisfy Eqs.~(\ref{eq:value}) and (\ref{eq:equiv}), and write $g=g_{\rm e}+g_{\rm o}$ with $g_{\rm e}(x)=[g(x)+ g(-x)]/2$ and $g_{\rm o}(x)=[g(x)- g(-x)]/2$. Then, for every kernel $f$, $f \text{ is chiral} \iff g_{\rm o}\neq 0 $,
and the mirror kernel is realized by the same even part and the opposite odd part, $\Phi[f^{\rm R}]=g_{\rm e}-g_{\rm o}$.

The proof is short. If $g$ is even, value matching gives $f(k)=g(x_k)=g(-x_k)=g(x_{q\ominus k})=f(q\ominus k)$, so $f$ is nonchiral. Conversely, if $f=f^{\rm R}$, equivariance gives $g(x)=\Phi[f^{\rm R}](x)=g(-x)$. The last statement is Eq.~(\ref{eq:equiv}) itself [SM, Sec.~S1.2].

The two implications of Theorem~1 rest on different hypotheses. ``Nonchiral $\Rightarrow g_{\rm o}=0$'' requires equivariance and is an instance of Curie's principle \cite{Curie2024,norton2016curie}: a rule that respects the reflection cannot turn a symmetric input into an asymmetric output. The principle, however, provides neither the converse nor a criterion for whether a construction respects the reflection. Here the converse follows from value matching alone, since $f$ can be read back from $g$ on the grid. Value matching is, up to a constant that does not enter $g_{\rm o}$, what reproducing the discrete energies requires, and it fixes the odd part on the grid, $g_{\rm o}(x_k)=f_{\rm o}(k)$, so that $g_{\rm o}$ carries the handedness of $f$ and not only its presence. An equivariant rule without value matching can erase chirality or reverse it [SM, Sec.~S1.3]. Equivariance, in turn, fails for some rules that reproduce all discrete energies (below) and holds necessarily for the minimal faithful realization (Theorem~2). The content of Theorem~1 is therefore the identification of the property (i.e., equivariance on top of value matching) that makes the odd part of a continuous realization an exact measure of discrete chirality.

\begin{figure}[t]
\includegraphics[width=0.95\columnwidth]{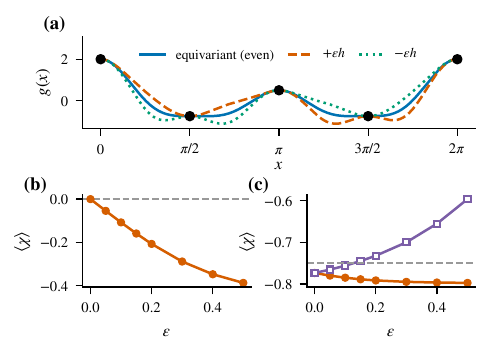}
\caption{Value matching does not fix the odd part.
(a) Nonchiral $\mathbb{Z}_4$ kernel $f_{\rm e}$ (dots) and realizations with its grid values and zero grid slopes: the minimal equivariant one (even) and the same with $\pm\epsilon h$ added.
(b) Spurious chirality: $\langle\chi\rangle$ of $f_{\rm e}$ under the non-equivariant rule $\Phi_{\min}+\epsilon h$.
(c) Extra chirality: $\langle\chi\rangle_{f^+}$ (circles) and $-\langle\chi\rangle_{f^-}$ (squares) under the same rule; equivariance requires them to coincide.
Dashed lines: discrete Boltzmann values; the offset at $\epsilon=0$ in (c) is a finite-temperature effect [SM, Sec.~S2].}
\label{fig:1}
\end{figure}

\textit{Two failure modes.}---Equivariance cannot be dropped, even when a realization reproduces the discrete energies exactly. To see this, consider the odd function
\begin{equation}
h(x)=\sin x+\tfrac12\sin 3x-\tfrac12\sin 5x ,
\end{equation}
which for $q=4$ vanishes together with its derivative at every grid point, $h(x_k)=h'(x_k)=0$. Let $\Phi_0$ be any rule that satisfies Eqs.~(\ref{eq:value}) and (\ref{eq:equiv}) and keeps every grid point stationary, $\Phi_0[f]'(x_k)=0$; the minimal such rule is constructed explicitly below [Eq.~(\ref{eq:minimal})]. Consider the rule $\Phi_\epsilon[f]=\Phi_0[f]+\epsilon h$ with $\epsilon$ fixed. It satisfies value matching and keeps every grid configuration stationary with its exact discrete energy, for every kernel and every graph. It is not equivariant, since $\Phi_\epsilon[f^{\rm R}](x)-\Phi_\epsilon[f](-x)=2\epsilon h(x)$, and it fails in two complementary ways [Fig.~\ref{fig:1}(a)]. A nonchiral kernel acquires an odd part $\epsilon h$ (i.e., a spurious chirality); and a chiral kernel and its mirror image acquire odd parts $g_{\rm o}+\epsilon h$ and $-g_{\rm o}+\epsilon h$, which are no longer opposite, so the two handednesses are realized with different strengths.

Because $h$ vanishes with its slope on the grid, neither failure shows in the energies or equilibria of discrete states; it appears through fluctuations. We sample $E$ with overdamped Langevin dynamics at temperature $T$, whose stationary density is $\propto e^{-E/T}$, map each phase to its nearest grid point, and measure the net chirality $\chi=N^{-1}\sum_i \sin(2\pi(s_i\ominus s_{i+1})/q)$ on a directed ring of $N=5$ spins ($J_{i,i+1}=-1$, $K=1$, $K_{\rm s}=3$, $T=0.3$), using an exact transfer matrix [SM, Sec.~S2]. For the nonchiral kernel $f_{\rm e}=(2,-\tfrac34,\tfrac12,-\tfrac34)$, $\langle\chi\rangle$ vanishes at $\epsilon=0$ but reaches $-0.39$ at $\epsilon=0.5$ [Fig.~\ref{fig:1}(b)]. For the chiral kernel $f^{+}=(2,\tfrac12,\tfrac12,-2)$ and its mirror $f^{-}=(f^{+})^{\rm R}$, the relation $\langle\chi\rangle_{f^-}=-\langle\chi\rangle_{f^+}$ holds only at $\epsilon=0$ [Fig.~\ref{fig:1}(c)]. Exact discrete energies therefore do not substitute for equivariance.

\textit{Faithful realizations.}---Theorem~1 does not single out a realization; a physical requirement does. A continuous realization is faithful to the discrete model if every discrete configuration is reproduced by its grid point $\boldsymbol{\theta}^\star=(x_{s_1},x_{s_2},\dots,x_{s_N})$, for every graph and every set of couplings. Faithfulness can be graded by how much of the landscape around the grid points is matched. At first order, which we use throughout, (F0) the continuous energy at $\boldsymbol{\theta}^\star$ equals $K H_f(\mathbf{s})$ up to an $\mathbf{s}$-independent constant, and (F1) $\boldsymbol{\theta}^\star$ is an equilibrium, $\nabla E(\boldsymbol{\theta}^\star)=0$. Higher orders also match the curvatures at the grid points, which matters only for finite-temperature statistics. First-order faithfulness guarantees that the continuous system has the same configurations, with the same energies, as the discrete model, independent of $K_{\rm s}$. Without it, grid configurations relax off the grid and lower their energies by state-dependent amounts, so that level crossings and degeneracies of the discrete model are shifted or lifted at any finite pinning strength. This spin softening, the physics of interest in Ref.~\cite{seno1994spin}, is a systematic error for a continuous realization of a given discrete model.

\textit{Proposition.}---A realization is first-order faithful if and only if
\begin{equation}
g(x_k)=f(k)+c, \quad g'(x_k)=0, \quad k\in\mathbb{Z}_q,
\label{eq:faithful}
\end{equation}
for some constant $c$. Sufficiency holds because the pinning term is constant and force-free on the grid; necessity follows from a single edge, whose two phases visit all differences $x_k$ [SM, Sec.~S1.4]. We refer to Eq.~(\ref{eq:faithful}) as the equilibrium-preserving conditions; that they reproduce the discrete level crossings at every pinning strength is verified in [SM, Sec.~S3]. Under Eq.~\eqref{eq:faithful}, the grid points are equilibria at every $K_{\rm s}$; they are strict local minima once $K_{\rm s}/K$ exceeds a threshold, and for larger $K_{\rm s}/K$ they are the only local minima [SM, Sec.~S1.4].

\textit{Theorem 2 (Unique minimal realization).}---Let
\begin{equation}
g(x)=\sum_{m=1}^{q}a_m\cos(mx)+\sum_{m=1}^{q-1}b_m\sin(mx).
\label{eq:minimal}
\end{equation}
For every kernel $f$, exactly one choice of the $2q-1$ coefficients satisfies
Eq.~(\ref{eq:faithful}) with $c=0$. No trigonometric polynomial of degree less than $q-1$
satisfies Eq.~(\ref{eq:faithful}) for every $f$, for any $c$ and with or without a constant term.

Substituting Eq.~(\ref{eq:minimal}) into Eq.~(\ref{eq:faithful}) gives $q$ value conditions and $q$ slope conditions that are linear in $(\mathbf{a},\mathbf{b})$. In this space the slopes at the grid points always sum to zero, so one slope condition is redundant, and the coefficients solve
\begin{equation}
M_q\begin{pmatrix}\mathbf{a}\\ \mathbf{b}\end{pmatrix}=\begin{pmatrix}\mathbf{f}\\ \mathbf{0}\end{pmatrix},
\label{eq:linear}
\end{equation}
where $M_q$ is the $(2q-1)\times(2q-1)$ matrix of the values and $q-1$ of the slopes of the basis functions at the grid points, $\mathbf{f}=(f(0),f(1),\dots,f(q-1))^{\rm T}$, and $\mathbf{0}$ has $q-1$ entries. The first part of Theorem 2 is the statement that $\det M_q\neq 0$ for every $q$; the proof and the explicit form of $M_q$ and of its solution of \eqref{eq:linear} are given in [SM, Sec.~S1.5].

The lowest degree is set by the slope conditions, not by the constant term:
below degree $q-1$ the first harmonic has no partner $1-q$ that aliases to it
on the grid, so its values and slopes there cannot be matched independently. At degree $q-1$ the solution is again unique: the term $a_q\cos(qx)$, with $a_q=F_0=q^{-1}\sum_k f(k)$, is dropped and the mean is carried by $c=-F_0$; the two forms have the same odd part and differ only in the curvature at the grid points, i.e., in the pinning needed to stabilize the grid [SM, Sec.~S1.5]. We use the form (\ref{eq:minimal}) throughout.

\textit{Corollary 1 (Minimal faithful realizations are equivariant).}---Reflection $g(x)\mapsto g(-x)$ maps the space (\ref{eq:minimal}) to itself, since it leaves the cosines unchanged and reverses the sines, and it maps a solution of Eq.~(\ref{eq:faithful}) for $f$ to a solution for $f^{\rm R}$, because $g(-x_k)=g(x_{q\ominus k})=f(q\ominus k)$ and the slopes remain zero. Uniqueness then forces $\Phi_{\min}[f^{\rm R}](x)=\Phi_{\min}[f](-x)$: the rule $\Phi_{\min}$ defined by Theorem~2 is reflection equivariant, without imposing equivariance as an extra condition. By Theorem~1, the minimal faithful realization encodes discrete chirality exactly: $f$ is chiral if and only if some $b_m\neq0$, and chiral reversal flips every $b_m$ while leaving every $a_m$ unchanged. These are not the discrete Fourier coefficients of $f$: for $f^+=(2,\frac{1}{2},\frac{1}{2},-2)$ the discrete Fourier transform gives a single sine coefficient, $\tfrac54$, at the first harmonic, whereas $\mathbf{b}=(\tfrac{15}{16},0,-\tfrac{5}{16})$, because the slope conditions split it over the first and third harmonics, which coincide on the grid. Beyond the minimal degree, faithfulness and equivariance become independent. The function $h$ of Fig.~\ref{fig:1} has degree $5>q$ and is invisible to Eq.~(\ref{eq:faithful}), so faithful realizations of higher degree need not be equivariant and can misrepresent chirality. Minimality is what ties the two requirements together.

\textit{Corollary 2 (Even couplings realize only nonchiral kernels).}---If a realization satisfies (F0) with an even pair function, $g(x)=g(-x)$, then $f(k)+c=g(x_k)=g(-x_k)=f(q\ominus k)+c$, so $f$ is nonchiral, whatever the degree of $g$. Oscillator Ising, clock, and Potts machines \cite{chou2019analog,mallick2022computational,cheng2026physics} use even couplings and therefore cannot represent a chiral kernel on the $N$ phases themselves; they must embed it in a larger spin space, e.g., by one-hot encoding each $\mathbb{Z}_q$ spin in $q$ Ising spins with a penalty for invalid states \cite{lucas2014ising}. The minimal realization represents it with one oscillator per spin.


\textit{Chirality as an orientation-odd phase shift.}---The gradient dynamics $\dot{\boldsymbol\theta}=-\nabla E$ of the minimal realization takes the form of a multiharmonic phase coupling,
\begin{equation}
\begin{split}
\dot\theta_i={}&-K\sum_{j}J_{ij}\sum_{m=1}^{q}R_m\sin\big(m(\theta_i-\theta_j)+\alpha_{m,ji}\big)\\
&-K_{\rm s}\sin(q\theta_i),
\end{split}
\label{eq:dyn}
\end{equation}
where the sum runs over the neighbors of $i$, $\alpha_{m,ji}$ is the phase shift of the channel that carries the influence of $j$ on $i$, $R_m=m\,(a_m^2+b_m^2)^{1/2}$, and $\alpha_m=\arg(a_m-ib_m)$. For an edge $(i,j)\in\mathcal{E}$ the two channels carry opposite shifts, $\alpha_{m,ji}=\alpha_m$ and $\alpha_{m,ij}=-\alpha_m$; the $q$th harmonic, which has no sine term in Eq.~(\ref{eq:minimal}), has $\alpha_q\in\{0,\pi\}$. The shift is thus orientation odd. By Corollary~1, chiral reversal maps $b_m\to-b_m$ at fixed $a_m$, i.e., $\alpha_m\to-\alpha_m$ at fixed $R_m$, which exchanges the shifts of the two channels of every edge. A kernel is nonchiral if and only if every $\alpha_m\in\{0,\pi\}$, i.e., the two channels coincide. The Sakaguchi lag \cite{sakaguchi1986soluble}, in contrast, is the same in both channels of an edge, i.e., orientation even. For two channels of equal amplitude with shifts $\beta_m\pm\delta_m$, the dynamics is a gradient flow if and only if $\beta_m\equiv0\pmod\pi$ [SM, Sec.~S4.2], so a Sakaguchi lag is non-variational and belongs to no realization of a kernel. The orientation-odd shift is likewise distinct from nonreciprocal couplings \cite{fruchart2021nonreciprocal}, which cannot be derived from a common pair potential: here the two channels of every edge derive from the same function $g$, the dynamics remain a gradient flow, and chirality is a property of the energy rather than a violation of action--reaction symmetry.

\begin{figure}[t]
\includegraphics[width=0.95\columnwidth]{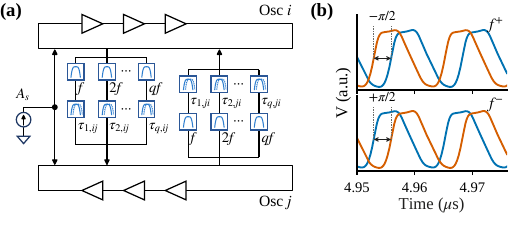}
\caption{Chirality as direction-dependent delays.
(a) Two oscillators coupled in each direction through band-pass filters at $f,2f,\dots,qf$ (solid symbols) and direction-dependent delays $\tau_{m,ji}$ or $\tau_{m,ij}$ (dashed symbols); a common $q$th-harmonic injection $A_{\rm s}\cos(2\pi qft-\gamma)$ pins both phases to the grid. Chiral reversal exchanges $\tau_{m,ji}\leftrightarrow\tau_{m,ij}$.
(b) Steady-state waveforms of two oscillators coupled by $f^+$ (top) and by $f^-$ (bottom), realized by exchanging the delays; the locked phase differences are $-88^\circ$ and $+88^\circ$.}
\label{fig:2}
\end{figure}


\textit{Direction-dependent delays.}---In a network of self-sustained oscillators, Eq.~(\ref{eq:dyn}) arises from phase reduction \cite{demir1998phase,levantino2012computing,nakao2016phase}. Each oscillator $j$ drives oscillator $i$ through separate channels for the harmonics $m=1,\dots,q$, selected by band-pass filters, and a $q$th-harmonic injection at frequency $qf$ provides the pinning term [Fig.~\ref{fig:2}(a)]. A delay $\tau$ in the $m$th channel shifts the phase of the transmitted harmonic by $-2\pi mf\tau$, so the phase shift $\alpha_{m,ji}$ is set by the channel delay
\begin{equation}
\tau_{m,ji}=\frac{\alpha_{m,ji}-\pi/2+\varphi_m}{2\pi mf}\pmod{1/mf},
\label{eq:delay}
\end{equation}
where $\varphi_m$ collects the phase of the filter and the coupling circuitry, which is the same for both orientations and independent of the kernel [SM, Sec.~S5]. Since chiral reversal maps $\alpha_{m,ji}\to\alpha_{m,ij}$, it exchanges the delays of the two channels of every edge, $\tau^{-}_{m,ji}=\tau^{+}_{m,ij}$, at fixed amplitudes. Delay-coupled oscillators \cite{yeung1999time,earl2003synchronization} usually have one transmission delay for all links and both directions, which in a locked state acts as an orientation-even lag; here the delays are direction-dependent and chosen per harmonic, and their antisymmetric part is the discrete chirality.

\begin{figure}[t]
\includegraphics[width=0.95\columnwidth]{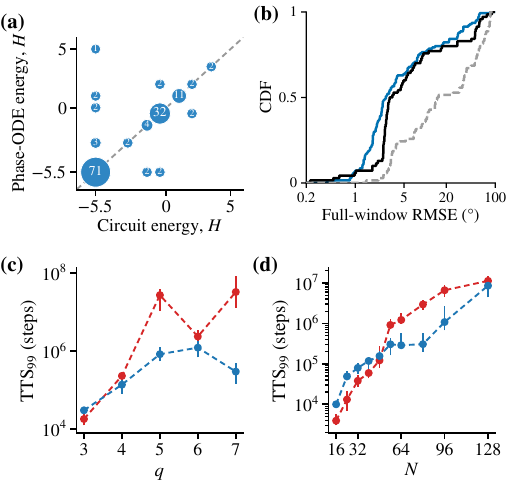}
\caption{(a,b) Mixed-level circuit simulation of a four-oscillator $\mathbb{Z}_4$ network programmed with $f^+$ (F) and with exchanged direction-dependent delays realizing $f^-$ (R).
(a) Final-state energy $H$, circuit vs phase model, for all 140 runs; numbers count runs per point.
(b) Cumulative distributions of the trajectory deviation: circuit vs phase model (blue), circuit F vs mirrored R (black), and F vs the mirrored prediction without delay exchange (dashed gray).
(c,d) Median TTS$_{99}$ over 40 test instances, with 95\% bootstrap confidence intervals, for the direct (blue) and one-hot (red) encodings, (c) versus $q$ at $N=48$ and (d) versus $N$ at $q=4$. Time steps are restricted to the range in which the integration resolves the dynamics; operation counts, simulated annealing, and unrestricted time steps are reported in [SM, Sec.~S7].}
\label{fig:3}
\end{figure}

\textit{Two oscillators.}---We test this in mixed-level circuit simulations (Cadence Spectre): transistor-level ring oscillators at $f\approx81$~MHz, behavioral band-pass filters, and delayed controlled sources for the couplings [SM, Secs.~S6.1 and S6.2]. For $q=4$, the kernel $f^+$ requires direction-dependent delays only in the first and third harmonics, where $b_m\neq0$; in the second and fourth the delays are equal in both directions. For a single edge with $J_{12}=-1$, the discrete ground state of $f^+=(2,\tfrac12,\tfrac12,-2)$ is $s_1\ominus s_2=3$, i.e., $\theta_1-\theta_2=-90^\circ$, and that of $f^-$ is $+90^\circ$. Two oscillators coupled by $f^+$ lock to $\theta_1-\theta_2=-88^\circ$, and exchanging the delays, i.e., realizing $f^-$, locks them to $+88^\circ$ [Fig.~\ref{fig:2}(b)]. From 729 initial conditions near in phase, which favor neither handedness, the two kernels lock to opposite phase differences [SM, Sec.~S6.3].

\textit{Four-oscillator network.}---We next simulate four oscillators on the graph with oriented edges $\mathcal{E}=\{(1,2),(2,3),(3,4),(1,4)\}$ and $J_{ij}=-1$, programmed with the same kernel $f^+$ (configuration F) and with the direction-dependent delays exchanged to realize $f^-$ (configuration R); the two netlists differ only in the delay values. We run 70 initial states, 64 of them close to discrete configurations, each in F and, with mirrored initial phases, in R (140 runs) [SM, Sec.~S6.4]. A phase model of the form (\ref{eq:dyn}), with one coefficient per channel computed from the measured waveform, PPV, filters, gains, and delays and frozen before the runs, predicts the circuit: the final discrete energies agree in 124 of 140 runs [Fig.~\ref{fig:3}(a)], and the final configurations themselves, which can differ between degenerate states, in 119; the median root-mean-square deviation between circuit and model phase trajectories over the full window is $2.7^\circ$ [Fig.~\ref{fig:3}(b), blue; SM, Sec.~S6.5].

The circuit also tests the mirror relation directly: if exchanging the delays realizes the chiral partner, then $\boldsymbol\theta_{\rm R}(t)=-\boldsymbol\theta_{\rm F}(t)$. The median deviation is $3.3^\circ$ [Fig.~\ref{fig:3}(b), black], close to the circuit--model level, compared with $15^\circ$ predicted without the exchange [Fig.~\ref{fig:3}(b), gray]; a circuit run without the exchange confirms this and ends in a metastable state [SM, Sec.~S6.6].

The mirror relation is not exact: 50 of the 64 grid-centered initial states lead to mirrored final configurations. The circuit couplings are not exactly reciprocal: at $m=1$ the two channels of an edge differ in amplitude by about $15\%$, owing to filter leakage of the high-gain harmonic channels [SM, Sec.~S6.6]. This non-variational part accounts for the violations: removing it from the calibrated phase model raises the mirrored outcomes from 50, as in the circuit, to 59 of 64.

\textit{Direct versus one-hot encoding.}---By Corollary~2, an oscillator machine with even couplings can represent a chiral kernel only by embedding it in a larger spin space; we compare the standard such embedding, one-hot encoding, with the direct realization. The instances are random 3-regular graphs with an intrinsically chiral kernel, random edge orientations, and a random $\mathbb{Z}_q$ gauge field [SM, Sec.~S7]. We anneal the phase dynamics (\ref{eq:dyn}) and an oscillator Ising machine with one-hot encoded spins \cite{wang2021solving,lucas2014ising} under the same protocol, tuned on separate training instances, and measure the number of integration steps needed to reach a certified ground state \cite{huangfu2018parallelizing} with probability $0.99$. At fixed $N=48$, the one-hot encoding needs fewer steps at $q=3$, whereas the direct encoding needs fewer at every $q\ge4$ [Fig.~\ref{fig:3}(c)]. At $q=4$, the one-hot encoding needs fewer steps for small $N$ and the direct encoding for larger $N$ [Fig.~\ref{fig:3}(d)]; counted in elementary operations, i.e., evaluations of coupling and pinning terms, the direct encoding is cheaper at every $N$ [SM, Sec.~S7.4]. In operation counts, simulated annealing \cite{kirkpatrick1983optimization} remains cheaper than both encodings simulated on a conventional processor; the time to solution of oscillator hardware is not addressed here [SM, Sec.~S7.5].


\textit{Conclusion.}---For any realization that matches the discrete values and treats mirror kernels alike, the chirality of a $\mathbb{Z}_q$ kernel is exactly the odd part of the continuous coupling, and the minimal faithful realization satisfies both conditions automatically. Dynamically, this odd part is an orientation-odd phase shift rather than a Sakaguchi lag; in oscillator networks it is the antisymmetric part of harmonic-resolved, direction-dependent delays, and in circuit simulations exchanging the delays reverses the handedness up to deviations set by non-reciprocal channel amplitudes. Because even couplings cannot carry chirality, the direct realization avoids the embedding that existing oscillator machines require. More broadly, algebraic handedness in a discrete spin Hamiltonian can be represented by tunable direction-dependent delays in continuous nonlinear dynamics, which opens a route to programmable physical simulators in which discrete symmetries, including chirality, are programmed by the delays.


\emph{Acknowledgments}--- This material is based upon work supported by the National Science Foundation under grant no. 2328961.


\bibliography{chiral_model}

\end{document}